\documentstyle[aps,prl,epsf,floats,twocolumn]{revtex}

\begin{document}


\twocolumn[\hsize\textwidth\columnwidth\hsize\csname@twocolumnfalse%
\endcsname

\def\r{{\bf r}}

\title{Oxygen impurities in NiAl: relaxation effects}
\author{David Djajaputra and Bernard R. Cooper}
\address{
Department of Physics, West Virginia University, PO BOX 6315, Morgantown, WV 26506, USA}

\date{\today}

\maketitle

\begin{abstract}
We have used a full-potential linear muffin-tin orbital method
to calculate the effects of oxygen impurities on the electronic structure of NiAl.
Using the supercell method with a 16-atom supercell we have investigated
the cases where an oxygen atom is substitutionally placed at either a nickel or an
aluminum site.
Full relaxation of the atoms within the supercell was allowed. We found
that oxygen prefers to occupy a nickel site over an aluminum site with a site selection
energy of 138 mRy (21,370 K). An oxygen atom placed at an aluminum site is found
to cause a substantial relaxation of its nickel neighbors {\it away}\/ from it. In
contrast, this steric repulsion is hardly present when the oxygen atom occupies the 
nickel site and is surrounded by aluminum neighbors. We comment on the possible 
relation of this effect to the pesting degradation phenomenon 
(essentially spontaneous disintegration in air) in nickel aluminides. 
\end{abstract}

\newpage

]


Nickel aluminides are potentially important industrial alloys in 
high-temperature structural applications, such as gas turbine engines, because 
they possess many desired physical properties. 
Among these are their low density, high strength, and good thermal conductivity, 
the last being an important factor in dissipating the heat generated by the engines.
The great interest in these alloys has resulted in a large body of literature.
\cite{darolia1991,deevi1997,miracle1993,westbrook1995,liu1995,miracle1995} 


Many properties of NiAl have been intensely investigated using various computational 
electronic structure methods. The band structure of NiAl has been experimentally studied
using photoemission and calculated using the linear augmented Slater-type orbital 
method by Lui {\it et al.} \cite{lui1990} They found that NiAl behaves like a 
good itinerant metal: the self-energy corrections in NiAl are significantly less 
than in pure ferromagnetic nickel. Band structure results should therefore provide a 
good description of the electronic properties of NiAl. 
Kim {\it et al.}\cite{kim1991} also calculated
the band structure of NiAl using a semirelativistic linearized augmented plane wave 
method and compared the result with the experimental optical spectra.
\cite{kim1991,schlemper1994} Their result
agreed with the result of Lui, and they found no need to incorporate self-energy
corrections into the spectrum in order to fit their experimental data. This is markedly
different from the result for CoAl where the corrections are needed to make a 
good fit.\cite{kim1991}
One can therefore put some confidence in the band-structure results for NiAl.


Most research on nickel aluminides has been devoted to the effort to understand 
the complex effects of impurities and other kinds of crystal defect.\cite{liu1997} Uses
of nickel aluminides, especially in polycrystalline form, as structural materials are 
limited by their brittleness at high temperature (over 800 $^\circ$C) where they become prone 
to brittle intergranular fracture and by their low ductility at intermediate temperatures.
Certain additive impurities have been found to greatly improve the room-temperature 
intergranular cohesion and the tensile ductility. The most notable of these cohesion 
enhancers is boron\cite{aoki1979} which can improve the tensile ductility by an order 
of magnitude and change the fracture mode from intergranular to transgranular.\cite{liu1997}  
Sun {\it et al.} \cite{sun1995} have studied the effects of boron and hydrogen on the 
cohesion of Ni$_3$Al using a full-potential linear muffin-tin orbital (FP-LMTO) method.
They concluded that boron improves the local cohesion by reducing the bonding charge
directionality around the nickel atoms and by inducing an increase of the interstitial
bonding charge.

\begin{figure}
      \epsfysize=45mm
      \centerline{\epsffile{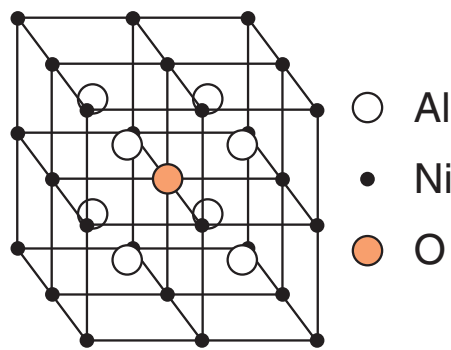}}
\bigskip
\caption{The 16-atom supercell used in the present work. The cubic supercell is built up
from $2^3$ unit cells of NiAl. The impurity atom (O) is placed at the center of the 
supercell, substituting for Ni in this picture (supercell Ni$_7$Al$_8$O). 
We also use a similar supercell, where the oxygen atom is placed at an Al site
(supercell Ni$_8$Al$_7$O).} 
\label{supercell}
\end{figure}

At intermediate to high temperature, nickel aluminides, and many other intermetallics,
exhibit brittle intergranular fracture due to the oxygen-induced embrittlement.\cite{liu1997} 
At high temperature, the oxidation will
selectively attack the least noble constituent, which is aluminum, and form the oxide 
product Al$_2$O$_3$.\cite{aitken1966} 
In some cases, these oxide products can provide a stable oxide layer that protects
the alloy underneath it from further oxygen attack. NiAl is one of the intermetallics
that are able to form a protective oxide layer and therefore is used as a coating for other 
intermetallic alloys.\cite{doychak1995}

The oxidation resistance decreases as we move toward the 
nickel-rich part of the Ni-Al phase diagram. The most damaging effect of oxygen
occurs when it causes the pesting degradation phenomenon (essentially
spontaneous disintegration in air). Some intermetallic
compounds that form protective coatings at high temperature literally
disintegrate when heated in the intermediate temperature range.\cite{doychak1995,aitken1966}
This occurs only on polycrystalline samples, so it is sometimes also
termed intergranular attack.\cite{doychak1995} 
Among nickel aluminides, pesting phenomenon has been 
observed in NiAl\cite{aitken1966,westbrook1964} and Ni$_3$Al.\cite{chuang1991}

Despite the extreme effects that it can cause, we are unaware of any previous
ab-initio study performed on the effects of oxygen in nickel
aluminides. Moreover, most studies of other impurities 
are limited by not allowing the atoms within the
supercell to relax. Relaxation effects are especially relevant to the case of impurity
atoms located at the grain boundary where they can be relatively free to move and 
bond to certain constituent atoms of the host alloy.


In this work we have used a full-potential linear muffin-tin orbital 
method within the local density approximation (LDA)
to study the effects of oxygen impurities on the electronic structure of NiAl. 
The details of this method have been documented elsewhere.
\cite{price1989,price1992,wills2000} For the work reported in this paper, 
we use nine ($spd$) orbitals for each atom in the supercell
and assign three LMTO basis functions for each orbital, corresponding to 
$\kappa$ = $-1.50$,$-0.50$, and $1.00$, respectively. Here $\kappa$ is the 
wave-number parameter; the square of $\kappa$ is the absolute value of the energy of the
basis function, measured relative to the muffin-tin zero, while its sign is equal to
the sign of the energy.\cite{skriver1984} The large number of basis
functions used provides our full-potential method with good flexibility to find the 
lowest-energy density.

\begin{table}
\caption{Equilibrium lattice constant $a_r$  for fully relaxed supercell,
its total energy $E_r$, and the corresponding relaxation energy $E_R$ 
(relative to the unrelaxed supercell).}
\vskip 5pt
\begin{tabular}{@{\hspace{10 pt}}cc@{\hspace{25 pt}}cc@{\hspace{10 pt}}cc@{\hspace{10 pt}}}
 Supercell & $a_r$ (a.u.) & $E_r$ (Ry) & $E_R$ (mRy) \\
\hline 
Ni$_8$Al$_7$O & 10.5878 & $-27,831.7387$ & $-0.675$  \\
Ni$_7$Al$_8$O & 10.6567 & $-25,278.8899$ & $-1.236$  \\
\end{tabular}
\label{energytable}
\end{table}


The 16-atom supercell that was used in the computation is shown in Fig.~\ref{supercell}.
Pure NiAl crystallizes in the B2 
structure with lattice constant of 5.4450 a.u.\cite{pearson1958} The calculated equilibrium
lattice constant using our FP-LMTO method is 5.3451 a.u which agrees with the experimental
value within 2\%. The supercell is constructed from $2^3$ unit cells of NiAl. The oxygen
impurity atom is placed at the center of the supercell, replacing a nickel or an aluminum
atom. For a given supercell lattice constant, we allow the atoms in the unit cell to 
relax to find the minimum total energy for that lattice constant. By symmetry, only the
eight nearest neighbors of the oxygen atom are allowed to relax, and they can only move 
radially away or toward the center oxygen atom (we do not consider the possibility of
symmetry breaking). Only one parameter is needed to describe the relaxation, namely the
distance of these neighboring atoms from the central oxygen atom.



\begin{figure}
      \epsfxsize=55mm
      \centerline{\epsffile{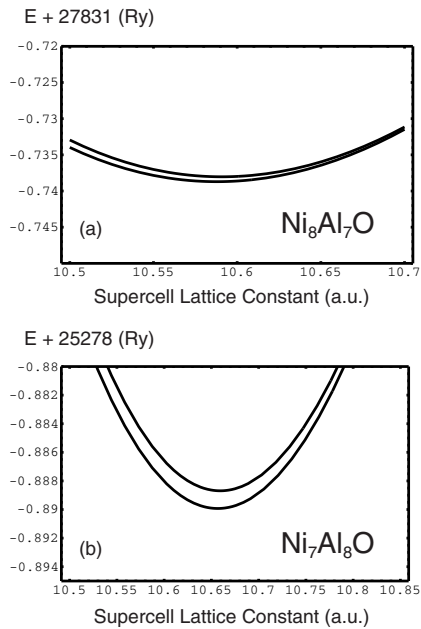}}
\bigskip
\caption{The FP-LMTO total energy with 16-atom supercell for the
case of oxygen substituting for an aluminum atom (a), 
and oxygen substituting for a nickel atom (b).
In both panels, the top curve is for an unrelaxed supercell while the bottom curve is
for the relaxed supercell. The equilibrium lattice spacing and the relaxation energy 
are listed in Table \ref{energytable}.} 
\label{energycurve}
\end{figure}

The computed total energies for the relaxed 16-atom supercell are shown in 
Fig.~\ref{energycurve} and compared with the curves for the unrelaxed supercell. 
Each continuous curve shown in Fig.~\ref{energycurve} is obtained by 
fitting 8--9 LMTO data points computed around the minimum-energy lattice constant using
an eight-parameter fitting function. The calculated relaxation energies $E_R$ are listed in
Table \ref{energytable}. We use the following definition for the relaxation energy

\begin{equation}
E_R = E_r(a_r) - E_u(a_u),
\end{equation}
with $a_u$ and $a_r$ being the lattice constant that give the minimum energy 
for the unrelaxed ($E_u$) and relaxed ($E_r$) configurations, respectively. 
The computed energy curves lie very close to each other, and the relaxation energies 
are relatively small, being about 100 K and 200 K (in equivalent temperature scale)
for oxygen at the Al and Ni sites, respectively. These numbers are about the same as 
the accuracy of the LMTO-LDA method, which has been estimated at about 1 mRy (158 K).
\cite{barth1994} Numerically, however, our FP-LMTO method did not produce any 
notable fluctuation and we believe that some confidence can be placed in these 
numbers.

\begin{figure}
\epsfysize=70mm
\centerline{\epsffile{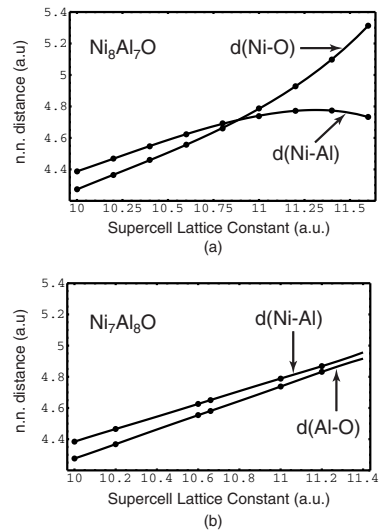}}
\bigskip
\caption{Nearest-neighbor (n.n.)
distances between atoms in the unit cell on the (011) plane as a function
of the supercell lattice constant. (a) Oxygen at Al site. For small lattice 
constants the movement of the Ni atoms is toward the central oxygen 
atom and away from the corner Al atoms. 
At large lattice constants the direction of movement reverses:
the Ni atoms are displaced significantly closer to the corner Al atoms.
This presumably reflects a steric repulsion between oxygen and nickel in the NiAl
environment. 
(b) Oxygen at Ni site. 
Apparently, the Al atoms are attracted about equally strongly by the O and the Ni atoms. 
This results in small relaxation.} 
\label{distances}
\end{figure}

The computed relaxation data are shown in Fig.~\ref{distances} where 
we track the nearest-neighbor distances between the atoms in 
the unit cell on the (011) plane as we change the supercell lattice constant. 
Figure \ref{distances}(b) displays the results for the case of oxygen at the Ni 
site. Shown are d(Al$-$O), which is the distance between the central O atom and one 
of its eight nearest-neighbor Al atoms, and d(Ni$-$Al), which is the distance between the
Al atom and its nearest neighbor Ni atom (at one of the corners of the supercell). 
Without relaxation, the oxygen atom will be at $(0,0,0)$, the aluminum atom at, e.g.,
$({1 \over 4},{1 \over 4},{1 \over 4})$, and the nickel atom at 
$({1 \over 2},{1 \over 2},{1 \over 2})$. As we vary the lattice constant $a$, the
distances scale linearly with it: d(Al$-$O) = d(Ni$-$Al) = $a \sqrt{3} /4$. This relation is 
still approximately followed when we relax the atoms, as shown in Fig.~\ref{distances},
with the aluminum atoms only attracted slightly more toward the central oxygen atom.

A significantly different situation occurs if we place the oxygen atom at the Al
site, as shown in Fig.~\ref{distances}(a). In this case, the nearest neighbors of the
central O atom are Ni atoms, the distance between them is d(Ni$-$O), and each of 
these Ni atoms neighbors an Al atom at its nearest corner of the supercell, where the distance
between them is denoted by d(Ni$-$Al). Under pressure, for lattice constants smaller
than the equilibrium value, the Ni atoms slightly relax toward the central oxygen atom, 
away from the corner Al atoms. However, as the lattice constant is 
increased above the equilibrium value, the intervening Ni atoms very quickly
start to move {\it away} from the central oxygen atom and relax  
closer to their neighboring Al atoms. We view this as reflecting a mutual 
repulsion between the oxygen and nickel atoms in the NiAl environment. 
Experimentally, the formation rate of aluminum oxide during exposure of nickel 
aluminides to oxygen is known to be much higher than that of nickel 
oxide.\cite{doychak1995,aitken1966}

The relaxation behavior of the atoms is thus seen to differ markedly depending on 
whether the impurity oxygen atom occupies an Al or a Ni site, especially in the 
stretched-supercell case, where the atoms have greater freedom to move around. This 
may have some relevance to the case of oxygen attack on polycrystalline nickel 
aluminides (pesting). We believe this reflects a situation where 
the oxygen atoms cause a sort of ``wedge effect'', 
seeping into the grain boundaries to reduce the intergranular cohesion and opening up 
the polycrystal wider for more infiltration of oxygen. This provides a {\it self-propelling} 
mechanism for oxygen atoms to infiltrate a polycrystal of nickel aluminide and, along the 
way, destroy the intergranular cohesion between the grains, effectively disintegrating the
polycrystal. In this scenario the pesting phenomenon can be seen to be fueled by the 
{\it combination} of two major factors: the thermal intergranular diffusion of oxygen and the
strongly {\it preferential} bonding of oxygen with one of the components of the alloy 
(aluminum in nickel aluminides). More insights into this phenomenon can perhaps be obtained by 
performing a molecular-dynamics (MD) simulation using microscopic parameters that are 
extracted from an ab-initio calculation such as reported in this paper. 
Campbell {\it et al.} have performed such an MD simulation for the oxidation of aluminum 
nanoclusters.\cite{campbell1999} 

The energetics of the preferential bonding of oxygen in NiAl can be studied 
by calculating the site selection energy.\cite{khowash1993} The total energy 
for pure bulk metal (per atom) is used to provide a reference energy for the constituent
species. In our case, these are the total energies of fcc Ni and fcc Al. These have
been calculated using the same FP-LMTO method and their values are

\begin{equation}
E({\rm Ni}) = -3,036.8304 \ {\rm Ry},
\end{equation}

\begin{equation}
E({\rm Al}) = -483.8440 \ {\rm Ry}.
\end{equation}
The total energies for the 16-atom supercell with one oxygen atom replacing a Ni or
an Al atom are listed in Table \ref{energytable}. 
We list the minimum of the total energy of the relaxed supercell.
The site selection energy is defined to be the difference 
between the following two values\cite{khowash1993}:

\begin{equation}
E_{\rm Ni} = E({\rm Ni}) + E({\rm Ni}_7{\rm Al}_8{\rm O})
 = -28,315.7204 \ {\rm Ry},
\end{equation}

\begin{equation}
E_{\rm Al} = E({\rm Al}) + E({\rm Ni}_8{\rm Al}_7{\rm O})
 = -28,315.5827 \ {\rm Ry}.
\end{equation}
Taking the difference, an oxygen atom in NiAl will prefer to occupy a nickel site with
the site selection energy of

\begin{equation}
\Delta E = 137.7 \ {\rm mRy} = 21,730 \ {\rm K}.
\end{equation}

That oxygen will prefer to occupy a nickel site over an aluminum site seems to contradict 
one's expectation based on the atomic radii of the constituent atoms. Since
the atomic radius of oxygen is closer to that of aluminum rather than nickel, one would
expect that the oxygen would prefer to occupy an aluminum site over the nickel site.
To understand our result, we need to recall the previously deduced mutual repulsion
between oxygen and nickel in the NiAl environment. An oxygen atom will therefore prefer
to be surrounded by nearest neighbors of aluminum, rather than nickel, atoms. It will
achieve this simply by occupying a nickel site.

\begin{figure}
      \epsfysize=90mm
      \centerline{\epsffile{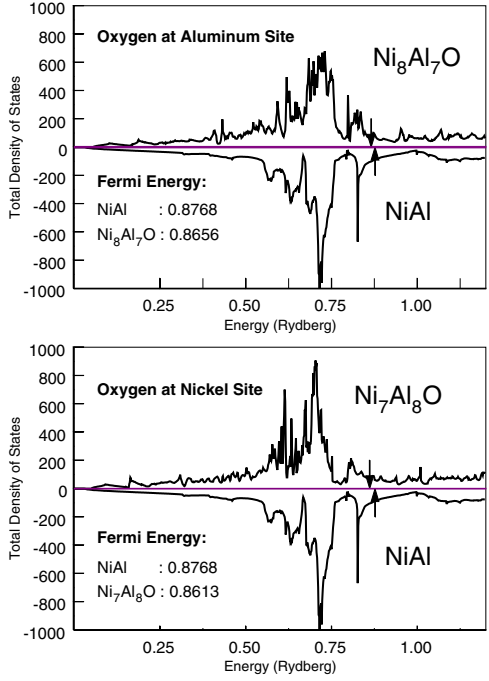}}
\bigskip
\caption{Total density of states for the 16-atom supercell calculated using the
FP-LMTO method. For comparison, in each panel we also show the negative of the 
density of states for pure NiAl. Vertical arrows on the
energy axis point to the positions of the Fermi energy. (a) Oxygen at Al site. 
(b) Oxygen at Ni site.} 
\label{totaldos}
\end{figure}

Finally, in Fig.~\ref{totaldos} we show the total density of states (DOS) for the two 
supercell
systems that we study in this work, Ni$_8$Al$_7$O and Ni$_7$Al$_8$O, calculated
from the FP-LMTO energy bands using the tetrahedron method. For comparison,
we also show the corresponding total DOS for pure NiAl. The dominant feature of the NiAl
DOS is the existence of the sharp peaks due to the $d$-orbitals of nickel, which  
hybridize only weakly with the other orbitals. Placing the oxygen atom at the Al site
in Ni$_8$Al$_7$O allows for some hybridization between these $d$-orbitals and the delocalized
$p$-orbitals of oxygen. This results in reduced sharpness of the Ni $d$-state peaks in 
the DOS without essentially any shift in the position of the peaks, as can be seen 
in Fig.~\ref{totaldos}(a). On the other hand,
if we place the oxygen impurity at the Ni site, then the nickel atoms will not have 
the oxygen atom as their nearest neighbor. The oxygen atom will have the aluminum atoms
as its nearest neighbors. Being the more electronegative element, oxygen will
interact with the valence electrons of Al and will localize a portion of those electrons
around itself. This depletes the Ni sites of some of the electrons from the Al that was 
formerly occupying its site in the pure NiAl case. The end result of this is a lower
electrostatic potential at the Ni sites and the reduction of the onsite energies 
of the $d$-orbitals of Ni, without much change in their spatial extent.
This translates to an almost rigid downward shift in the position of the Ni $d$-state
peaks in the DOS without much alteration in their width. This downward shift of about 25 mRy 
can quite readily be discerned in Fig.~\ref{totaldos}(b).

\smallskip

\noindent {\it Acknowledgements}. This work benefited from much discussion with our late
colleague David L. Price. A large part of the computational work 
for this project was performed at the Maui High Performance Computing 
Center (MHPCC). This work was supported by 
AF-OSR Grant F49620-99-1-0274.

\end{document}